\documentclass[twocolumn,preprintnumbers,amsmath,amssymb]{revtex4-1}
\usepackage{graphicx}
\usepackage{dcolumn}
\usepackage{bm}
\usepackage{braket}
\usepackage{adjustbox}

\newcommand{\comment}[1]{}
\newcommand{\yb}{$^{174}$Yb }
\newcommand{\g}{^{1}S_{0}}
\newcommand{\tpl}{{^3P}_{1}}

\newcommand{\br}{\langle}
\newcommand{\ke}{\rangle}

\newcommand{\ddel}{\ensuremath{\dot{\delta}}}
\newcommand{\omrec}{\omega_{\rm rec}}

\usepackage{color}

\begin{document}
	
	\title{Excited-Band Bloch Oscillations for Precision Atom Interferometry}
	
	\author{Katherine E. McAlpine, Daniel Gochnauer, and Subhadeep Gupta}
	
	\affiliation{Department of Physics, University of Washington, Seattle WA 98195}
	
	\date{\today}
	\begin{abstract}
	We propose and demonstrate a method to increase the momentum separation between the arms of an atom interferometer and thus its area and measurement precision, by using Bloch oscillations (BOs) in an excited band of a pulsed optical standing wave lattice. Using excited bands allows us to operate at particular ``magic'' depths, where high momentum transfer efficiency ($>99.4\%$ per $\hbar k$, where $\hbar k$ is the photon momentum) is maintained while minimizing the lattice-induced phase fluctuations ($<1\,$milliradian per $\hbar k$) that are unavoidable in ground-band BOs. We apply this method to demonstrate interferometry with up to 40$\hbar k$ momenta supplied by BOs. We discuss extensions of this technique to larger momentum transfer and adaptations towards metrological applications of atom interferometry such as a measurement of the fine-structure constant.  
	\end{abstract}
	\maketitle	

\section{Introduction} Bloch oscillations (BOs) describe the periodic motion of a particle in a lattice responding to a constant force. While this behavior emerged as a fascinating prediction of the Landau-Zener theory of electron conduction in an ionic lattice in the presence of an external electric field \cite{bloc28,zene34}, BOs remained only a theoretical construct until first observations in semiconductor superlattices \cite{wasc93}. Clean observations of BOs in neutral systems soon after, using ultracold atoms in an optical lattice \cite{bend96,wilk96,peik97}, provided an early benchmark for the use of trapped atomic gases as quantum simulators for condensed matter systems. Following this early work, BOs have been adapted into the atom optics community as a high-efficiency momentum transfer tool \cite{batt04,mcdo13,page19,gebb19} and have been fruitfully utilized in metrological applications ranging from testing quantum electrodynamics \cite{bouc11,park18}, to measuring local gravity \cite{poli11} and to testing the equivalence principle \cite{tara14}. 

The benefit of high efficiency in transferring momentum, which has been instrumental in ground-band BO based atom-optics applications, comes at the cost of uncontrolled phase shifts on the atomic wavefunction due to fluctuations in the lattice potential strength. This has limited the use of BOs as beam splitters within phase-stable atom interferometers (AIs) to relatively low momenta in earlier free-space geometries \cite{clad09,mull09,footkem3}. 

In this paper, we propose and demonstrate excited-band Bloch oscillations within an atom interferometer as a new tool for precision measurement which simultaneously exhibits high efficiency and low lattice-induced phase noise. Our proposal is based on the observation that unlike in the case of the ground band, BOs in an excited band when performed at a particular ``magic'' depth become relatively immune to lattice strength fluctuations. We experimentally establish our magic depth hypothesis for multiple excited bands by examining BO-induced phase shifts on one arm of a Mach-Zehnder interferometer operated on a Bose-Einstein condensate atom source. Our implementation is capable of high efficiency momentum transfer ($>99.4\%$ per $\hbar k$ where $2k$ is the lattice wavenumber) at these regions, allowing for interference signals with 40$\hbar k$ momenta supplied by BOs to one AI arm. Our results point to significant potential improvements in precision interferometric measurements of the fine structure constant $\alpha$ and a concurrent test of quantum electrodynamics, and can also impact other AI applications such as gravity measurements.

\begin{figure}
	\centering
	\includegraphics[width=1\linewidth]{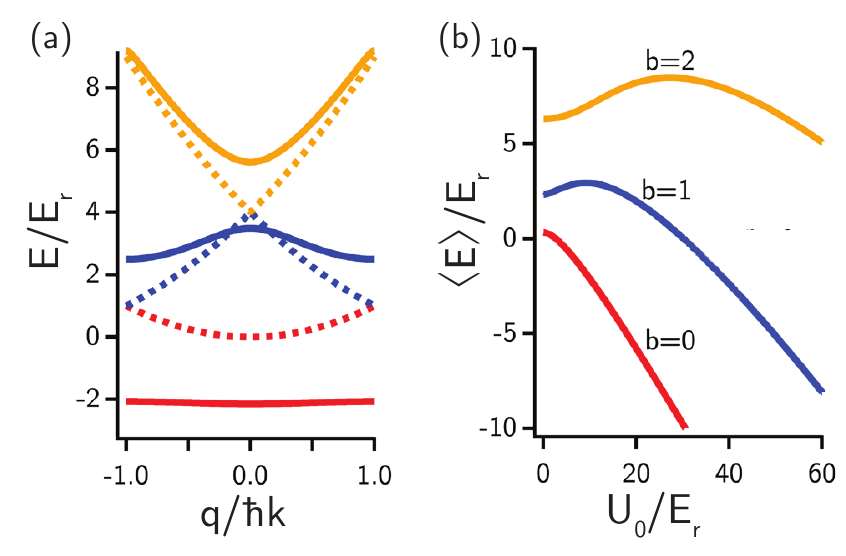}
	\caption{(a) Bloch bands (solid lines) for a sinusoidal optical lattice with a representative depth of $U_0$ = $10\,E_r$, where $E_r$ is the recoil energy. The atomic energies are computed by diagonalizing the Hamiltonian with 51 states from $\ket{-50\hbar k}$ to $\ket{+50\hbar k}$ in steps of $2\hbar$k, where $\hbar$k is the photon momentum. Dotted lines represent the quadratic free-space dispersion. (b) The average energy over one Brillouin-zone (from $q=-\hbar k$ to $q=+\hbar k$)  of the ground and first two excited bands $\langle E \rangle$. The magic depth for each excited band is at its respective local extremum. The ground band does not exhibit any magic depth feature.}
	\label{fig:fig1}
\end{figure}

We first explain our magic depth hypothesis by considering the lattice-induced phase shift in a Brillouin-zone picture where BOs correspond to the periodic oscillation of the quasimomentum as it gets Bragg-reflected at the zone boundary (Fig.\,\ref{fig:fig1}(a)). Such a Bloch-band picture is also useful for the analysis of Bragg diffraction processes from pulsed lattices in atom optics applications \cite{goch19}. For the linear rate of change of quasimomentum $q$ relevant to BOs, it is useful to consider the average $\langle E \rangle$ of the energy $E$ for a particular band taken over one Brillouin-zone (from $q=-\hbar k$ to $q=+\hbar k$). These are shown for the first three bands in Fig.\,\ref{fig:fig1}(b). The average energy $\langle E \rangle$ of the ground band is always negative and results in the atomic phase being strongly sensitive to unavoidable intensity fluctuations of the optical lattice. This behavior can be understood as arising from level repulsion in second-order perturbation from higher-lying bands. On the other hand, every excited band is repelled by bands both above and below. The quadratic free-space dispersion leads to a positive shift of $\langle E \rangle$ at low lattice depths, since the adjacent lower band is closer in energy than the adjacent higher one. At high lattice depths, these energy separations start to become comparable and the larger number of higher energy bands results in a negative shift of $\langle E \rangle$. A key observation from the Brillouin zone picture that is central to this work is that the average energy in an excited band must feature one local maximum as a function of lattice depth. When operated at this magic lattice depth, excited-band BOs feature phase shifts on the atomic wavefunction that are first-order insensitive to lattice-induced Stark shifts.

The rest of this paper is organized as follows. After discussing relevant technical details of the experiment in Section II, we demonstrate the differences between ground- and excited-band BOs in Section III. In Section IV we demonstrate the magic depth property of excited-band BOs using interferometry in a Mach-Zehnder geometry. We discuss our implementation of large momentum transfer (up to $40\hbar k$) within an AI using magic depth excited-band BOs in Section V, and also investigate their scalability towards larger numbers of BOs. Finally, we provide a summary and outlook for future applications in Section VI.

\section{Atom Source and Lattice Beams}
\label{sec:atomsandlattice}
Similar to earlier work \cite{plot18,goch19}, we perform our experiments with \yb Bose-Einstein condensates (BECs) consisting of $10^5$ atoms, formed by evaporative cooling in a $532\,$nm optical dipole trap. After BEC production, the atoms are released from the trap and allowed to expand for $3\,$ms before the first application of a pulsed optical lattice. 

To create the optical lattices for our BOs and other diffraction pulses, we use two counterpropagating horizontally-oriented laser beams which are detuned by $\Delta$ from the intercombination transition $(\g \rightarrow \tpl)$ at $556\,$nm which has linewidth $\Gamma=2\pi \times 182\,$kHz. For all experiments reported in this work, $\Delta/\Gamma=+3500$ or $+1300$, and its value is noted with each data set presented. The lattice beams have a waist of $1.8\,$mm, which is large compared to the size of the atom cloud ($<30\,\mu$m). A small (sub-MHz) relative frequency $\delta$ between the two lattice beams is controlled at the sub-Hz level using Direct-Digital Synthesizers as the radiofrequency sources to drive the corresponding acousto-optic modulators. 

Two kinds of pulsed lattices are used in this work. In one, a trapezoidal temporal intensity profile is used for the BO pulses in which the lattice is first turned on with an increasing intensity ramp to the desired depth $U_0$. The value of $\delta$ during this turn-on process is chosen to place atoms into desired quasimomenta in the lattice-frame. A frequency sweep $\ddel$ at fixed depth $U_0$ then provides the external force for the BO, following which the lattice is turned off with a decreasing intensity ramp. Atoms undergo Bloch oscillation with period $T_{\rm BO}=8\omrec/\ddel$ during the frequency sweep where $\hbar \omrec=E_r=\hbar^2k^2/(2m)$ is the recoil energy and $m$ is the mass of the atom. In order to load stationary atoms into a particular band, the initial relative detuning is chosen as $\delta=(b+0.5) 4\omrec$ where $b$ is the band number ($b=0$ is the ground band). This ensures lattice loading at quasimomentum $q=+(-)0.5 \hbar k$, for $b=$odd(even), away from band gaps to avoid interband transitions. To load moving atoms into a particular band, the initial $\delta$ is adjusted to meet this condition in the atom-frame. The timescale of the two intensity ramps are equal and chosen to always satisfy the adiabaticity criterion $\frac{1}{U_0}|\frac{\partial U_0}{\partial t}| \ll |\Delta E|/\hbar $ where $\Delta E$ is the energy separation from the eigenstate nearest to the state of interest. 

The second kind of pulsed lattices in this work are Bragg diffraction pulses with Gaussian temporal profiles characterized by rise and fall $1/e$ times $\simeq 30\,\mu$s \cite{goch19}. For these pulses, $\delta$ is kept at a constant value of  $4 N_B \omrec$ to match the resonance condition for an $N_B^{\rm th}$ order Bragg process \cite{goch19}. Example intensity profiles of both kinds of pulsed lattices can be found in Fig.\ref{fig:fig3}(a). 

\section{Bloch Oscillations in Ground and Excited Bands}
\label{sec:BOge}
\begin{figure}
	\centering
	\includegraphics[width=1\linewidth]{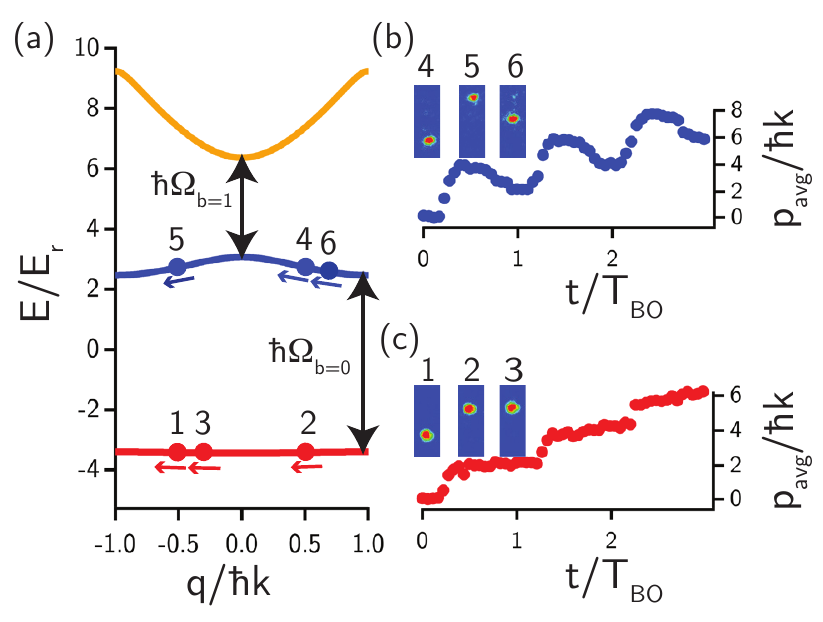}
	\caption{(a) Ground- and first-excited band Bloch state trajectories during a BO at $U_0$ = $13.6 E_r$. For the ground band, the atom starts at 1, traverses the avoided crossing at $q=-1\hbar k$, gains $2\hbar k$ in lab-frame momentum, and continues to move at that momentum for the duration of the BO (2 and 3). For the excited-band BO trajectory shown (4-6), an atom's lab-frame momentum increases by $4\hbar k$ as it moves past the $q=0$ avoided crossing from 4 to 5, then decreases by $2\hbar k$ at the $q=-1\hbar k$ avoided crossing, such that it has a net lab-frame momentum gain of $2\hbar k$ by the time it reaches 6. The mean atomic momentum is plotted as a function of sweep time (corresponding to a particular final detuning for a fixed $\ddel$) for an atom undergoing a partial sweep through the Brillouin zone in (b) the first-excited band and (c) the ground band. Insets in (b) and (c)  show representative time-of-flight absorption images. $\Delta/\Gamma$ is $+3500$ and the frequency sweep $\ddel$ is 2$\pi \times 83\,$kHz/msec for all the data. The intensity ramp times are $300\,\mu$s for (b) and $600\,\mu$s for (c).}
	\label{fig:fig2}
\end{figure}

To illuminate the differences between ground- and excited-band BOs, we first present how the atomic momentum evolves during such BO processes. The frequency sweep $\ddel$ corresponds to the atom changing its quasimomentum in the lattice frame. In its quasimomentum trajectory, as an atom approaches and moves past an avoided crossing while remaining in the same band, its momentum in the lab frame changes in even units of $\hbar k$ (see Fig.\,\ref{fig:fig2}(a)). Thus, an atom can be accelerated to large values in the lab frame using several cycles of BOs.

In the case of ground-band BOs, lab-frame acceleration occurs only when an atom traverses the avoided crossing at a Brillouin zone boundary at $q=\mp 1\hbar k$, where the quasimomentum changes from $\mp 1\hbar k$ to $\pm 1\hbar k$. In an excited band, in addition to the $q=\mp 1\hbar k$ crossing, there is another avoided crossing at $q=0$. Thus, in excited-band BOs, an atom changes its lab-frame momentum twice during one Bloch period. However, the total change in the lab-frame momentum is always $\pm[2(b+1)\hbar k - 2b\hbar k] = \pm2\hbar k$ for one Bloch oscillation. 

This behavior can be seen in Fig.\,\ref{fig:fig2}(b,c) where the lab-frame momentum during Bloch oscillations in the ground- and first-excited band are shown for frequency sweeps with varying final quasimomenta. The final momentum distribution was measured using time-of-flight absorption imaging. The average momentum computed from these images shows the differences between the ground- and excited-band BOs described above. 

For applications in AI, a high efficiency of Bloch oscillation is desirable, which in turn provides high-efficiency momentum transfer to an atomic wavepacket. As the quasimomentum is swept using $\ddel$, atoms can tunnel to other bands at the locations of avoided crossings, potentially making the BO process inefficient. The Landau-Zener model gives this tunneling probability as
\begin{equation}
P_{\rm LZ} = {\rm exp}\left(-\frac{\pi \Omega^2}{2\beta \ddel}\right),
\label{eqn:LZ}
\end{equation} 
where $\hbar \Omega$ is the band gap at the avoided crossing and $\beta$ is the higher of the two band numbers participating in the avoided crossing \cite{zene32,heck02}. $P_{\rm LZ}$ increases when the band gap is reduced or when the quasimomentum is swept faster. To successfully perform a high-efficiency BO, $P_{\rm LZ}$ must be small. Even though there are two avoided crossings for each excited band, the one at $q=0(\pm \hbar k)$ will always be smaller for $b$ odd(even) and will have a greater contribution to tunneling loss during a BO. For the parameters of the data in Fig.\,\ref{fig:fig2}, $P_{\rm LZ} \simeq 10^{-8} (10^{-25})$ for the loss probability during one BO in the first-excited (ground) band.

\section{Magic Depth Bloch Oscillations in Excited Bands}
\label{sec:MD}
\begin{figure}
	\centering
	\includegraphics[width=1\linewidth]{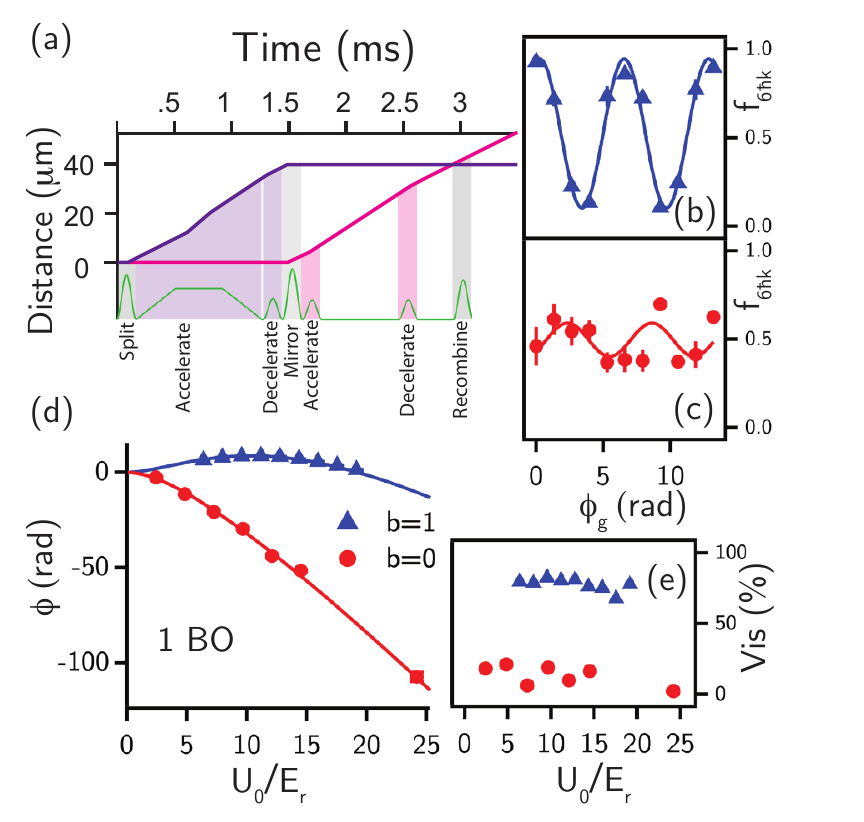}
	\caption{(a) Space-time diagram of the Mach-Zehnder interferometer (to scale) showing the upper and lower arms (purple and pink lines respectively) and atom-optics sequence (green line) for one BO in either the ground or an excited band applied to the upper arm. The purple (pink) highlight indicates that a particular pulse affects the momentum of the upper (lower) arm. Pulses which affect the momentum of both arms are highlighted in gray. (b) Representative interferometer signal for one BO in the first-excited band ($U_0=11.2\,E_r$). Each data point is the average of three population measurements. Solid line is the best-fit sinusoid. (c) Same as (b) but for $b=0$ and $U_0=9.7\,E_r$. (d) The interferometer phase shift as a function of lattice depth for one BO in the ground band (red filled circles) or first-excited band (blue filled triangles) shows good agreement with Bloch-bands calculations (solid lines) for $\Phi_{\rm BO}$ (Eq.\,\ref{eqn:phid}). (e) Visibilities of the interferometer signals for $b=1$ and $b=0$. The BO parameters are $\Delta/\Gamma=+1300$, $\ddel=2 \pi \times 83\,$kHz/msec and intensity ramp times of $300\,\mu$s for all the data in this figure.}
	\label{fig:fig3}
\end{figure}

To verify the magic depth hypothesis and explore its applicability in interferometry, we systematically applied BOs in one arm of a Mach-Zehnder interferometer (MZ) and analyzed the resultant phase shifts (Fig.\,\ref{fig:fig3}). The basic beamsplitter-mirror-recombiner pulses of our MZ consist of $\pi/2-\pi-\pi/2$ third-order Bragg pulses coupling states $|0\ke$ and $|6\hbar k\ke$.

We first compare the effects of phase shifts from a BO in $b=0$ with that in $b=1$ within the MZ. The BO pulse is applied on one arm (upper one in Fig.\,\ref{fig:fig3}(a)) during the first half of the MZ, inmediately following the beamsplitter pulse, accelerating the upper arm from the state $\ket{6\hbar k}$ to the state $\ket{(n+6)\hbar k}$. After a short free evolution time ($10\,\mu$s in Fig.\,\ref{fig:fig3}(a)), this arm is then decelerated back to $|6\hbar k\rangle$ with a Bragg $\pi-$pulse of order $n/2$ ($n=2$ in the figure). The mirror pulse then swaps the momenta of the upper and lower arms. In the second half of the MZ, the $\ket{6\hbar k}$ (lower) arm is accelerated to $\ket{(n+6)\hbar k}$ using a Bragg $\pi-$pulse, and is allowed to evolve for some time ($630\,\mu$s in Fig.\,\ref{fig:fig3}(a)) before being decelerated back to $\ket{6\hbar k}$ with another Bragg $\pi-$pulse. A final $\pi/2$ recombiner pulse is applied when the two arms overlap and the population in each momentum state is measured using time-of-flight absorption imaging. As the phase $\phi_{\rm g}$ of the recombiner pulse is varied, the fractional population $f_{6\hbar k}$ in the $\ket{6\hbar k}$ state oscillates sinusoidally (see Fig.\,\ref{fig:fig3}(b,c)). The phase $\Phi$ of this oscillation is the difference in the phases accrued by the two interferometer arms. We determine this MZ phase by fitting a sinusoid function and observing its behavior as a function of the BO pulse depth $U_0$ for the ground and excited bands (see Fig.\,\ref{fig:fig3}(d)).

The depth-dependence of the MZ phase is well explained using our Bloch-bands model. During the BO pulse, each interferometer arm is in a particular band of the lattice. As the depth $U_{0}$ is varied, the band energies change, resulting in a phase shift on each interferometer arm given by
\begin{equation}
\frac{1}{\hbar}\int_{\rm pulse}(E(q(t),U_0(t))-E_f(q(t)))dt
\label{eqn:dphase}
\end{equation}
which is the difference between the band energy $E(q(t),U_0(t))$ and the free-space energy $E_f(q(t))$, where $U_0(t)$ and $q(t)$ are determined by the parameters of the BO pulse. During the BO pulses, the phase accrual for each arm can be organized into contributions from the intensity ramps and the frequency sweep. Since the final MZ phase is the difference between phase accrued by each arm, the interferometer phase shift due to a BO pulse can be written as:
\begin{equation}
\Phi_{\rm BO} = \phi_{1,I}+\phi_{1,f}-\phi_{2, I}-\phi_{2, f}
\label{eqn:phid}
\end{equation}
where $\phi_{i,I}$ is the phase accrued by arm $i$ (where $i=1(2)$ is the upper(lower) arm in Fig.\,\ref{fig:fig3}(a)) during the intensity ramp and $\phi_{i,f}$ is the phase accrued by arm $i$ during the frequency sweep. Each of these phase shifts is calculated using Expression \eqref{eqn:dphase}. 

It is important to note that the non-accelerating arm during the BO process (lower arm in Fig.\,\ref{fig:fig3}(a)) can make a non-negligble contribution in the evaluation of $\Phi_{\rm BO}$. However, since this arm is always in a higher band determined by its speed relative to the lattice frame, it performs Landau-Zener tunneling transitions across relatively small band gaps ($P_{\rm LZ} \simeq 1$) at each avoided crossing during the BO. Thus the lab-frame momentum of this interferometer arm does not change during the frequency sweep.

Figure \ref{fig:fig3}(d) demonstrates that in the first-excited band ($b=1$), there is a local maximum corresponding to a magic depth at $U_{0}\simeq 10 E_r$, while the ground band ($b=0$) exhibits monotonic behavior. Our observations are well-matched by our calculations based on the Bloch-bands picture. 

In addition to demonstrating the existence of the magic depth, Fig.\,\ref{fig:fig3} also shows the usefulness of this property for AI, as evidenced in the signal quality for $b=1$ versus $b=0$ (Fig.\ref{fig:fig3}(b,c)). We define the visibility of such signals as ${\rm Vis}=\frac{{\rm Max-Min}}{{\rm Max+Min}}\times 100\%$, where Max and Min refer to the maximum and minimum of the fitted sinusoid. As shown in Fig.\,\ref{fig:fig3}(e), the visibility for $b=1$ is around $80\%$, dramatically better than the $10\%$ level for $b=0$ over the entire range of depths explored. As another measure of the difference in signal stability and corresponding applicability for interferometry, the error in the fitted phase $\delta \Phi$ is 0.073 radians for $b=1$ and 1.4 radians for $b=0$, averaged over the presented data sets. These observations clearly show the high sensitivity of interferometer phase noise to lattice intensity fluctuations in ground-band BO. Furthermore, the visibility for the $b=0$ data trend downward with increasing depth (Fig.\,\ref{fig:fig3}(e)), consistent with an increased value of $U_0|\frac{\partial \br E\ke}{\partial U_0}|$ (Fig.\,\ref{fig:fig1}(b), Fig.\,\ref{fig:fig3}(d)) leading to greater phase noise.

\begin{figure}
	\centering
	\includegraphics[width=1\linewidth]{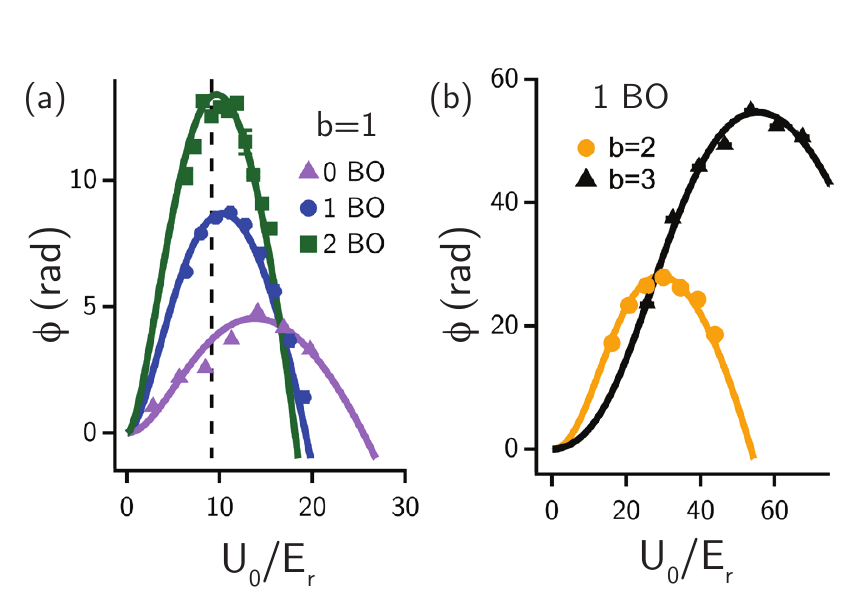}
	\caption{Magic depths observed as maxima in Mach-Zehnder interferometer phase shifts as functions of depth for excited-band BOs. (a) Shows data for the first-excited band for zero (purple triangles), one (blue circles), and two (green squares) BOs together with the Bloch-bands calculations (solid lines). Note that the ``1 BO'' data is the same as in Fig.\,\ref{fig:fig3}(d). The dashed line marks $U_{\rm MD}$ (see text). (b) Interferometer phases for BOs performed in the second (yellow) and third (black) excited band. The BO parameters are $\Delta/\Gamma=+1300$, $\ddel=2 \pi \times 83\,$kHz/msec and intensity ramp times of $300\,\mu$s for all the data in this figure.}
	\label{fig:fig4}
\end{figure}

The magic depth property is ubiquitous to excited-band BOs and we demonstrate its existence for different numbers of Bloch oscillations as well as for different excited bands. In Fig.\,\ref{fig:fig4}(a) we present measurements of depth-dependent MZ phase shifts for different numbers of BOs performed in the first-excited band. For these measurements, an interferometer geometry similar to Fig. \ref{fig:fig3}(a) was used with either 0, 1, or 2 BOs (corresponding to $n=0,2,4$) performed in $b=1$ in the first half of the MZ, using different ranges for the frequency sweep. In the 0 BO ($n=0$) case, only the intensity ramps were applied during the BO pulse on the upper arm in the first half of the MZ and no acceleration or deceleration pulses were applied on the lower arm during the second half. As can be seen in the figure, there is very good agreement between our theoretical model and experimental observations.

We define $U_{\rm MD}$ as the value of $U_0$ at which the condition $\frac{\partial \br E\ke}{\partial U_0}=0$ is satisfied. Thus $U_{\rm MD}$ for a particular band is independent of experimental parameters, as shown in Fig.\,\ref{fig:fig1}(b). In actual experiments however, the intensity ramps necessary to maintain adiabaticity lead to depth values for the condition $\frac{\partial \Phi_{\rm BO}}{\partial U_0}=0$ to be higher than $U_{\rm MD}$. This can be seen in Fig.\,\ref{fig:fig4}(a) as the position of the local maximum approaching the dashed line marking $U_{\rm MD}$ for $b=1$, as the number of applied BOs increases. Thus $U_{\rm MD}$ corresponds to the limiting value of the magic depth as the number of BOs increases and the intensity ramp time becomes insignificant compared to the frequency sweep time. 

Another feature of the plots in Fig.\,\ref{fig:fig4}(a) is that they cross each other at $U_0\simeq 17\,E_r$. This corresponds to the depth where the average band energy is equal to the average free-space ($U_0=0$) energy (see Fig.\ref{fig:fig1}(b)), and $\Phi_{\rm BO}$ in Eq.\,\ref{eqn:phid} only has contributions from the intensity ramps, which are the same for all the data sets. 

In Fig.\,\ref{fig:fig4}(b), we present measurements of depth-dependent MZ phase shifts for higher bands ($b=2$ and $b=3$). For these measurements, an interferometer geometry identical to Fig.\,\ref{fig:fig3}(a) was used and only the initial $\delta$ of the BO pulse was adjusted according to the desired band number. Our data clearly indicates that the magic depth increases with band number, and again we find very good agreement between our theoretical model and experimental observations. 

\section{Large Momentum Separation Interferometers with Excited-band Bloch Oscillations}
\label{sec:largemom}
To fruitfully apply the magic depth property towards creating large momentum separation between interferometer arms, additional criteria have to be considered. High frequency sweep rates are desirable to apply large momentum transfer in a short time. Furthermore, the decoherence from spontaneous scattering can start to play a role for large numbers of BOs. In this section we first demonstrate, in subsection A, the role of spontaneous scattering in determining optimum sweep rates for BOs. In subsection B we investigate how the magic-depth BO operation parameters scale with band number $b$ to guide future efforts in using this tool towards high precision interferometry.      

\subsection{Optimum Frequency Sweep Rate for high momentum transfer}

\begin{figure}
	\centering
	\includegraphics[width=1\linewidth]{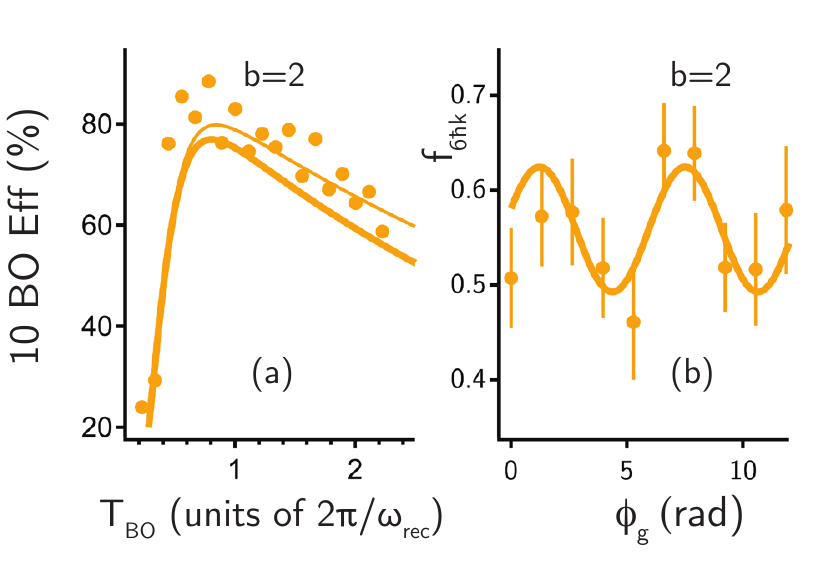}
	\caption{(a) The efficiency for 10 BOs in $b=2$ as a function of the BO period for $U_{0}=27.5 E_r$. The peak efficiency per $\hbar k$ of momentum gain was measured to be $99.4\%$. The thick and the thin solid lines are theoretical model curves (see text). (b) Interferometer signal for a sequence consisting of 20 BOs in band 2 with $U_{0}=27.5 E_r$ and $T_{\rm BO}=120\,\mu$s. Solid line is the best-fit sinusoid. $\Delta/\Gamma=+3500$ and intensity ramp times of $300\,\mu$s for all the data in this figure.} 
	\label{fig:fig5}
\end{figure}

The period of the Bloch oscillation, $T_{\rm BO}$ (equivalently $\ddel$), has to be chosen to optimize efficiency, balancing the considerations that a short $T_{\rm BO}$ (large $\ddel$) causes more tunneling to other bands (Eq.\ref{eqn:LZ}) while a long $T_{\rm BO}$ (small $\ddel$) causes more spontaneous scattering. We demonstrate this behavior by determining the efficiency of BOs as the fraction of atoms in the target momentum state in time-of-flight absorption images. The measured efficiency for a 10 BO ($20\hbar k$) pulse at $U_0=U_{\rm MD}$ in $b=2$ as a function of $T_{\rm BO}$ (Fig. \ref{fig:fig5}(a)) exhibits a clear maximum. The general features of our data are well-reproduced by a simple model (thick solid line in Fig.\,\ref{fig:fig5}(a)) incorporating $P_{\rm LZ}$ and spontaneous scattering in an expression for the efficiency of $n/2$ BOs: 
\begin{equation}
(1-P_{\rm int})^2 \times [(1-P_{\rm LZ})\times{\rm exp}(-R_s T_{\rm BO})]^{(n/2)}
\label{eqn:BOeffmodel}
\end{equation}
Here $P_{\rm int}$ is the spontaneous scattering probability during each of the two identical intensity ramps and $R_s=\frac{U_0}{8 \hbar}\frac{\Gamma}{\Delta}$ corresponds to the spontaneous scattering rate \cite{footkem1} during the frequency sweep, and we have taken the average intensity experienced by an atom in the standing wave to be one half of the peak intensity. This parameter-free model reproduces our observations quite well, exhibiting a maximum at an optimum sweep time which we label as $T_{\rm BO,opt}$.

We note that while this model reproduces the observed location of optimum efficiency, it slightly underestimates the value for the peak, which is observed to be $99.4\%$ per $\hbar k$. We can improve on the model by noting that the atomic wavefunction during the BO process is not uniform but localized at the antinodes of the blue-detuned lattice, which reduces the $R_s$ factor in Eq.\,\ref{eqn:BOeffmodel}. Accounting for the reduced average intensity due to the non-uniform spatial wavefunction produces the thin solid line shown in Fig.\,\ref{fig:fig5}(a), in better agreement with the data but still slightly underestimating the peak efficiency. 

In our theoretical calculations of $P_{\rm LZ}$ we have used the avoided crossing value at the minimum bandgap point, given by $q=\pm \hbar k$ for $b=2$. At $U_0=U_{\rm MD}$, the band dispersion still has substantial curvature (see Fig.\ref{fig:fig1}(a) where $U_0 \simeq U_{\rm MD}$ for $b=1$) and the use of Eq.\,\ref{eqn:LZ} should be reasonably valid. 
 
By operating magic-depth BOs near the optimum time determined in Fig.\,\ref{fig:fig5}(a) and extending the overall MZ time, we demonstrate stable interferometry with the application of 20 BOs ($40\hbar k$) on the upper arm only. This experiment was performed in a geometry similar to Fig.\,\ref{fig:fig3}(a) with $n=20$, but with a decelerating BO pulse replacing the Bragg deceleration pulse in the upper arm, allowing for a greater number of BOs to be applied. Even though the second BO pulse reverses the momentum transfer from the first BO pulse (with $20 \hbar k$ each), the lattice-induced phase shifts from both BOs have the same sign and add. As shown in Fig.\,\ref{fig:fig5}(b), we obtain a clear MZ signal for these conditions. The sinusoidal fit returns a visibility of $13\%$ and a phase error $\delta \Phi = 0.57\,$rad. Our current level of light intensity fluctuations of $<2\%$ \cite{plot18}, should contribute  $<40\,$mrad to the phase uncertainty (arising from the local curvature of $\langle E \rangle$ at $U_{\rm MD}$ and detailed below) \cite{footkem2}. The lower visibility and higher phase noise observed in Fig.\,\ref{fig:fig5}(b) compared to that in Fig.\,\ref{fig:fig3}(b) can be explained by noting that the total interferometer time for the 20 BOs experiment was $6.7\,$ms, which is more than twice as long as all other experiments reported in this work. The MZ geometry is sensitive to mirror vibrations which become a significant source of phase fluctuations at long interferometer times \cite{tory00}. We can thus infer that vibration noise is the principal contributor to the visibility reduction at these long interferometer times, consistent with other reports for vibration-sensitive interferometers \cite{tory00,jami14}.  

While Fig.\ref{fig:fig5}(b) demonstrates that excited-band BOs can be used to impart large momentum transfer within an interferometer, its full exploitation with the benefits of large $n$ values will require using an interferometer geometry that is insensitive to mirror vibrations, such as a contrast interferometer \cite{gupt02, jami14} or a simultaneous conjugate interferometer \cite{chio09}. 

\subsection{Scaling of Magic Depth Properties with Band Number}

In order to further assess the usefulness of the magic depth properties towards high-precision interferometry, we analyze how these properties scale with band number. Our results can serve as a guide in choosing experimental parameters to simultaneously optimize for high efficiency and suppression of lattice-induced phase noise. 

\begin{figure}
	\centering
	\includegraphics[width=1\linewidth]{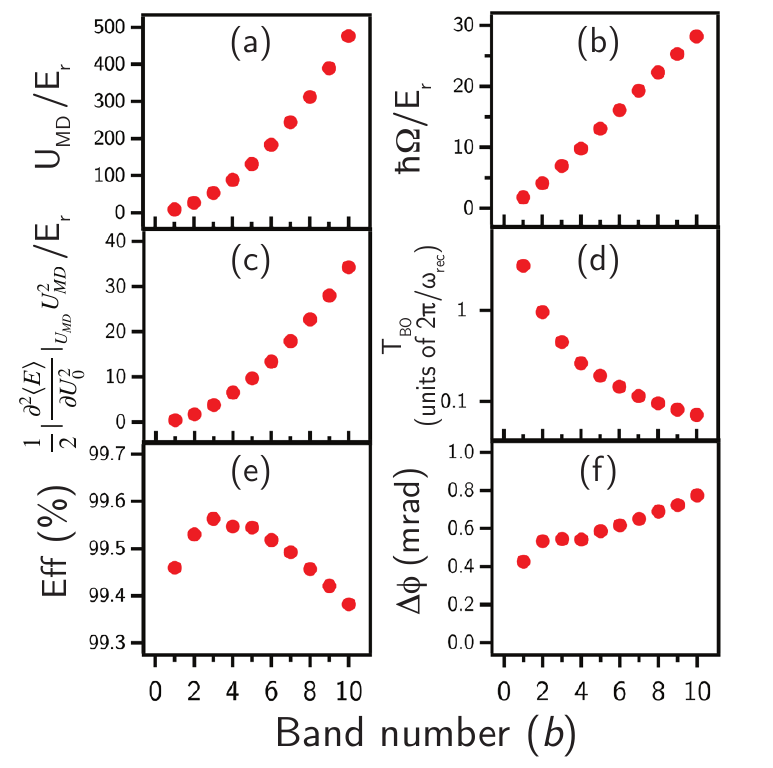}
	\caption{(a), (b), and (c) show calculated magic depth, bandgaps, and the quantity $\frac{1}{2}\lvert\frac{\partial^2 \langle E \rangle}{\partial U_{0}^2}\rvert_{_{U_{\rm MD}}}U_{MD}^2$ for the first ten excited bands. (d) shows $T_{\rm BO,opt}$ and (e) shows peak efficiency per $\hbar k$ for $\Delta/\Gamma=10^4$. (f) shows the calculated interferometer phase noise per $\hbar k$ for $1\%$ intensity noise.} 
	\label{fig:fig6}
\end{figure}

Calculated $U_{\rm MD}$ values and the corresponding (minimum) band gaps $\hbar \Omega$ for excited-bands up to $b=10$ are shown in Figure \ref{fig:fig6}(a,b). Both $U_{\rm MD}$ and $\hbar\Omega$ have the expected monotonically increasing behavior and show approximately quadratic and linear scaling with band number, respectively. Even when operating at the magic depth, residual interferometric phase fluctuations will arise from the local curvature of $\langle E \rangle$ at $U_{\rm MD}$. Since experimental intensity fluctuations are usually a fixed percentage of the average intensity, we plot $\frac{1}{2}\lvert\frac{\partial^2 \langle E \rangle}{\partial U_{0}^2}\rvert_{_{U_{\rm MD}}} \times U_{\rm MD}^2$ in (Fig. \ref{fig:fig6}(c)) to elucidate this scaling. The magic-depth properties shown in Fig. \ref{fig:fig6}(a-c) are common to all systems (provided sinusoidal lattices are used) and can be used to guide choices for experimental paramaters.

As shown in Fig.\,\ref{fig:fig5}(a), optimum BO efficiency requires choosing an optimum BO frequency sweep rate, which in turn depends on the choice of detuning $\Delta$. Since our Eq.\,\ref{eqn:BOeffmodel} (thick solid line in Fig.\ref{fig:fig5}(a)) already captures our observations quite well, we use this approach to evaluate the optimum sweep time $T_{\rm BO,opt}$ for a chosen $\Delta/\Gamma=10^4$ (see Fig.\ref{fig:fig6}(d)). The corresponding efficiency, which we take to be a conservative estimate, remains quite high at $\simeq 99.5\%$ per BO (Fig.\ref{fig:fig6}(e)). The expected phase fluctuations from lattice-induced shifts at $U_{\rm MD}$ can then be calculated as  $\Delta \Phi = 2\pi \times \frac{1}{2}\lvert\frac{\partial^2 \langle E \rangle}{\partial U_{0}^2}\rvert_{_{U_{\rm MD}}}\times U_{\rm MD}^2 \times \frac{1}{2}T_{\rm BO,opt} \times (\frac{r}{100})^2$ per $\hbar k$ for a relative intensity noise of $r\%$. The resulting plot of $\Delta \Phi$ as a function of $b$ (Fig.\ref{fig:fig6}(f), for $r=1$) shows that the expected phase noise grows very weakly with $b$ and stays below $1\,$mrad per $\hbar k$ for $b=1$ through 10. While this may suggest that the choice of $b$ is not very important, the strong dependence of $T_{\rm BO,opt}$ on $b$ favors using higher $b$ to minimize the time for acceleration processes and thus allow for longer free evolution time or interaction time for interferometric sensing. Finally, we note that the optimum choice for $b$ for a given application will also depend on the experimentally accessible maximum intensity for the lattice beams (see Fig.\ref{fig:fig6}(a)). 

\section{Outlook and Summary}

We have proposed and demonstrated the existence of particular lattice depths in excited-band Bloch oscillations where the average energy of the band is first-order insensitive to lattice intensity fluctuations. Operation at such magic depths can be maintained in combination with high efficiency of BOs, making this property of value to precision atom interferometry. Using experimentally confirmed operational characteristics, we have provided projections for scaling up of this method to very large momentum transfers. In particular, efficient momentum transfer with several hundred BOs seems feasible with manageable ($\lesssim 1$ radian) lattice-induced phase fluctuation effects. Our results are presented scaled to recoil energy and recoil momentum, and should be directly adaptable to interferometry with other atomic species. 

We can assess the improvements that this method can bring to a fine-structure constant measurement using contrast interferometry \cite{jami14}. In recent work, Bragg pulses were employed for acceleration to an inter-arm momentum separation of 112 photon recoils \cite{plot18} within such an interferometer. However, the performance was limited by the efficiency of momentum transfer, resulting in signal reduction for large momenta. Comparing the momentum transfer efficiency of 98.45\% per $\hbar k$ in \cite{plot18} with the highest values in Fig.\ref{fig:fig6}(e), we can conservatively expect a four-fold increase in momentum separation with magic-depth excited-state BOs, leading to a sixteen-fold improvement in sensitivity to $\alpha$ for the same interferometer time. By operating at $T_{\rm BO,opt}$ for $b=4$ to reduce the acceleration time, and in a vertical geometry to increase the free evolution time, an overall improvement of more than a factor of 100 is attainable, allowing for a sub-part-per-billion measurement of the fine structure constant and consequent test of QED theory. Our excited-band magic-depth BO technique may also benefit other AI applications that currently rely on Bragg diffraction processes for large momentum transfer including gravimetry \cite{mazz15} and gravity gradiometry \cite{asen17}.

\section*{Acknowledgments}

We thank Tahiyat Rahman for experimental assistance, Benjamin Plotkin-Swing for early work on the apparatus, and Katherine McCormick for a careful reading of the manuscript. This work was supported by NSF Grant No. PHY-1707575.


\begin{thebibliography}{29}
\bibitem[{foo(2019{\natexlab{a}})}]{footkem3}
\bibinfo{journal}{Ground-band BOs have been used to demonstrate a phase-stable
  $80\hbar k$ AI in a guided (non free-space) geometry \cite{mcdo13}. Recent
  work in free-space AIs has shown even larger momentum splitting with
  ground-band BOs \cite{page19,gebb19}, however phase-stability was
  demonstrated only up to $20\hbar k$.}
\bibitem[{foo(2019{\natexlab{b}})}]{footkem1}
\bibinfo{journal}{For our parameters, the ratio of the intensity to the
  saturation intensity is more than 3 orders of magnitude smaller than the
  quantity $(2\Delta/\Gamma)^2$.}
\end{thebibliography}
\end{document}